\documentclass[acmsmall,screen,authorversion,nonacm]{acmart} 
\usepackage{xspace}
\usepackage{graphicx}
\graphicspath{{figures/}}
\usepackage{needspace}

\usepackage{paralist}
\usepackage{ifthen}
\usepackage[normalem]{ulem} 
\usepackage{xcolor}

\newboolean{showedits}
\setboolean{showedits}{true} 
\ifthenelse{\boolean{showedits}}
{
	\newcommand{\del}[1]{\textcolor{red}{\sout{#1}}} 
	\newcommand{\nbe}[3]{
		{\colorbox{#3}{\bfseries\sffamily\scriptsize\textcolor{white}{#1}}}
		{\textcolor{#3}{\sf\small$\blacktriangleright$\textit{#2}$\blacktriangleleft$}}}
}{
	\newcommand{\del}[1]{} 
	
	\newcommand{\nbe}[3]{}
}
\usepackage[most]{tcolorbox}
\ifthenelse{\boolean{showedits}}
{
  \newtcolorbox{inserted}{%
       title=Inserted text:,
       colframe=blue,colback=blue!5!white,
       breakable,
       leftrule=0mm, 
       bottomrule=0mm,
       rightrule=0mm,
       toprule=0mm,
       arc=0mm, outer arc=0mm,
       oversize
  }
  \newtcolorbox{deleted}{%
       title=Deleted text:,
       colframe=red,colback=red!5!white,
       breakable,
       leftrule=0mm, 
       bottomrule=0mm,
       rightrule=0mm,
       toprule=0mm,
       arc=0mm, outer arc=0mm,
       oversize
  }
  \newtcolorbox{refactored}{%
       title=Rewritten text:,
       colframe=blue,colback=red!5!white,
       breakable,
       leftrule=0mm, 
       bottomrule=0mm,
       rightrule=0mm,
       toprule=0mm,
       arc=0mm, outer arc=0mm,
       oversize
  }
}{

}
\newboolean{showcomments}
\setboolean{showcomments}{true}
\newcommand{\id}[1]{$-$Id: scgPaper.tex 32478 2010-04-29 09:11:32Z oscar $-$}

\ifthenelse{\boolean{showcomments}}
{\newcommand{\nbc}[3]{
 {\colorbox{#3}{\bfseries\sffamily\scriptsize\textcolor{white}{#1}}}
 {\textcolor{#3}{\sf\small$\blacktriangleright$\textit{#2}$\blacktriangleleft$}}}
 }
{\newcommand{\nbc}[3]{}
 }

\newboolean{isblinded}
\setboolean{isblinded}{true}
\ifthenelse{\boolean{isblinded}}
{\newcommand\blind[1]{BLINDED\xspace}}
{\newcommand\blind[1]{#1\xspace}}

\newcommand{\ie}{\emph{i.e.},\xspace}
\newcommand{\eg}{\emph{e.g.},\xspace}

\newcommand{\sep}{\mbox{$\gg$}}
\usepackage[english]{babel}
\usepackage{listings}
\lstdefinelanguage{Smalltalk}{
  morestring=[d]',
  morecomment=[s]{"}{"},
  alsoletter={\#:},
  escapechar={!},
  literate=
    {BANG}{!}1
    {UNDERSCORE}{\_}1
    {_}{{$\leftarrow$}}1
    {>>>}{{\sep}}1
    {^}{{$\uparrow$}}1
    {~}{{$\sim$}}1
    {-}{{\sf -\hspace{-0.13em}-}}1  
    {+}{\raisebox{0.08ex}{+}}1		
    {-->}{{\quad$\longrightarrow$\quad}}3
	, 
  tabsize=4
}[keywords,comments,strings]

\definecolor{source}{gray}{0.95}

\newcommand{\st}{\lstinline[mathescape=false,backgroundcolor=\color{white},basicstyle={\sffamily\upshape}]}
\newcommand{\lst}[1]{{\textsf{\textup{#1}}}}
\lstnewenvironment{code}{%
	\lstset{%
		frame=single,
		framerule=0pt,
		mathescape=false
	}
}{}

\usepackage{subfigure}
\newboolean{preprint}
\setboolean{preprint}{true}
\ifthenelse{\boolean{preprint}}{
}{
}
\AtBeginDocument{%
  }
\copyrightyear{2024}
\acmConference[LIVE 2024]{LIVE 2024}{Oct.\ 20-25, 2024}{Pasadena, CA}

\usepackage{caption}
\newboolean{anonymous}
\setboolean{anonymous}{true}
\newcommand\anonymize[2]{\ifthenelse{\boolean{anonymous}}{#2}{#1}\xspace}

\begin{document}
\title[Example-driven development: bridging tests and documentation]{Example-driven development: \\ bridging tests and documentation}%
\ifthenelse{\boolean{preprint}}{%
\thanks{Presented at \href{https://2024.splashcon.org/home/live-2024}{Live 2024}, colocated with SPLASH 2024, Pasadena, USA, Oct 20-25, 2024.}%
}{}

\author{Oscar Nierstrasz}
\affiliation{%
  \institution{feenk gmbh}
  \city{Wabern}
  \country{Switzerland}}
\email{oscar.nierstrasz@feenk.com}

\author{Andrei Chi\c{s}}
\affiliation{%
  \institution{feenk gmbh}
  \city{Wabern}
  \country{Switzerland}}
\email{andrei.chis@feenk.com}

\author{Tudor G\^irba}
\affiliation{%
  \institution{feenk gmbh}
  \city{Wabern}
  \country{Switzerland}}
\email{tudor.girba@feenk.com}

\begin{abstract}
Software systems should be \emph{explainable}, that is, they should help us to answer questions while exploring, developing or using them.
Textual documentation is a very weak form of explanation, since it is not causally connected to the code, so easily gets out of date.
\emph{Tests}, on the other hand, are causally connected to code, but they are also a weak form of explanation.
Although some tests encode interesting scenarios that answer certain questions about how the system works, most tests don't make interesting reading.

\emph{Examples} are tests that are also factories for interesting system entities.
Instead of simply succeeding or failing, an example returns the object under test so that it can be inspected, or reused to compose further tests.
An example \emph{is} causally connected to the system, is always live and tested, and can be embedded into live documentation.
Although technically examples constitute just a small change to the way that to test methods work, their impact is potentially ground-breaking.
We show
\begin{inparaenum}[(i)]
	\item how Example-Driven Development (EDD) enriches TDD with live programming,
	\item how examples can be \emph{molded} with tiny tools to answer analysis questions, and
    \item how examples can be embedded within live documentation to make a system explainable.
\end{inparaenum}
\end{abstract}



\maketitle

\section{Background: Examples = Tests + Factories}\label{sec:intro}

Unit tests, as originally introduced by Beck \cite{Beck94c}, exercise objects, referred to as \emph{fixtures}, and evaluate assertions over these objects.
Fixtures are created with the help of shared \emph{setup} methods, and, once the test method succeeds or fails, are simply discarded.
Only if a test fails do we have access to the fixture, from within the debugger.

Gaelli noted that there was a missed opportunity here, and proposed a modified approach to unit testing in which tests return their fixtures, that is, they serve as factories for \emph{examples}~\cite{Gael06b}.
The output of a test --- an example --- can then be (re-)used as the input (fixture) for another test.
Examples can then be composed to form higher-level scenarios~\cite{Gael07a}.

In principle, any XUnit testing framework, for language X, can be extended to support examples.
JExample\footnote{\href{https://web.archive.org/web/20230925063519/https://scg.unibe.ch/research/jexample}{https://scg.unibe.ch/research/jexample}} extends JUnit 4, allowing test methods not only to return examples, but also to accept as input one or more other examples from the same test example class.
Interestingly, refactoring tests as examples establishes a partial order amongst test methods.
The impact of this is that
\begin{inparaenum}[(i)]
	\item code duplication is reduced since common preambles to complex tests are refactored as shared examples, and
	\item defect localization is improved since fewer tests will fail~\cite{Kuhn08a}.
\end{inparaenum}
H{\"a}nsenberger showed that tests in the wild often contain much duplicated code.
By performing dynamic analysis on tests, one can largely automate the process of migrating classical unit tests to examples~\cite{Haen08b,Haen09a}.

Although examples are already interesting as a means to make explicit the otherwise implicit dependencies between tests to reduce duplicated code and improve defect localization, it turns out that examples can impact the software process in other, important ways.

The key contributions of this paper are highlighted in the following sections.
In \autoref{sec:edd} we show how EDD enriches the TDD process with live programming opportunities.
Then, in \autoref{sec:moldable}, we show how EDD enables the \emph{moldable development} of tiny analysis tools.
In \autoref{sec:explainable} we show how EDD offers a means to augment documentation with live, interactive examples, as a step towards making software systems explainable.
We outline in \autoref{sec:applying} how EDD can be applied in other programming languages and environments.
Finally we discuss in \autoref{sec:related} how EDD builds on prior work, and conclude briefly in \autoref{sec:conclusion}.

%
%

\section{EDD: TDD + Examples}\label{sec:edd}

In this section we first motivate the need for extended TDD with examples, then we introduce EDD and \emph{Glamorous Toolkit} a development environment in which examples form a cornerstone, and finally we illustrate EDD with the help of a running example of modeling \emph{Prices} of goods in multiple currencies.

\subsection{Motivation}

A key motivation for integrating unit tests into mainstream software development was to enable continuous refactoring of evolving software systems \cite{Beck00a}.
Test-Driven Development (TDD)~\cite{Beck03a} offers a way not only to ensure that tests are produced in tandem with the implementation of a software system, but importantly to exploit the potential for tests to serve as requirements that can drive the design and implementation process.

Ignoring ongoing ideological debates over the pros and cons of TDD and other Agile practices, we can observe the following issues with TDD:
\begin{enumerate}[(i)]
	\item TDD advocates that one should always start development by writing a test first.
\emph{But how do you know how to express what you want to test?}
Sometimes you know up front how to create the setup and what the desired outcome should be.
But quite often you do not know what the right assertion should be, so you have to ``guess first'' to write the assertion.
Even when you do know the assertion up front, you still have to ``guess first'' to imagine an API to exercise the object and express the assertion.
Of course, this is how tests ``drive'' development, but could there be an easier way to get to the test code that you need to write?
	\item TDD then recommends that you write just enough code to make the test pass.
\emph{But how do you know where to start coding?}
Again you have to ``guess first'' to figure out where to begin coding.
	\item Green tests tell you that what worked before still works.
\emph{But what can you do with a green test except read the code?}
A red test is useful, as long as it brings you to a debugger that lets you explore why the test has failed, but a green test is uninteresting.
How could we make it useful?
\end{enumerate}

\subsection{EDD and GT}

\emph{Example-Driven Development} (EDD) offers a new take on TDD in which examples drive the development process.
It is similar to TDD, but can differ in important ways by exploiting opportunities raised by live programming.
We can summarize EDD as follows:
\begin{enumerate}[(i)]
	\item Instead of starting by writing a test, we \emph{start by inspecting a bare-bones example,} and incrementally write code that will become a test scenario to produce an interesting example.
	\item Instead of writing code to make a test pass, we \emph{iteratively and incrementally grow the example}, and extract assertions that express what is interesting about the example
	\item Instead of returning nothing, \emph{a green test returns an example} that we can inspect, interact with, and, as we shall see, embed into live documentation.
\end{enumerate}



Glamorous Toolkit\footnote{\url{https://gtoolkit.com}} (GT) is an open-source ``moldable development environment'' implemented on top of the Pharo\footnote{\url{https://pharo.org}} Smalltalk  platform.
The purpose of GT is to enable the creation of \emph{explainable software systems} that can answer both technical and business domain answers about the software with the help of cheaply-built custom tools.
The process of \emph{molding} the software is that of augmenting software entities with these custom tools.
Examples play an important part in the molding process in that
\begin{inparaenum}[(i)]
	\item GT is extensively covered by test cases, each of which returns an example,
	\item molding is an interactive process that typically starts from a live example, and
	\item live examples are used extensively to document GT itself in the \emph{GT Book}, a knowledge base of notebook pages containing embedded examples.
\end{inparaenum}


\subsection{Illustrating EDD with a Running Example}

We now illustrate the principles behind EDD with a running example of modeling multi-currency prices for goods, inspired by the currency example introduced by Beck and Gamma to explain unit testing~\cite{Beck98a}.
Although the prices example is rather artificial, it is small enough to illustrate the key points.
We provide some more realistic examples in \autoref{sec:explainable}.

%

Suppose we have an existing library of classes implementing amounts of \st{Money} in various currencies, such as \st{100 euros} or \st{10 usd}.
We would now like to implement \emph{prices} for goods, where a \st{Price} may be a concrete, fixed price, or a \emph{discounted} price, where the discount may be a fixed value or a percentage.

With TDD we would probably start by specifying a test method within a dedicated \st{TestPrice} class that sets up a fixture for a concrete fixed price of, say, \st{100 euros}, and then asserts something about it.
With examples, \emph{we could do pretty much the same}, creating an example method in a \lst{PriceExamples} class that creates the same fixture, asserts something, \emph{and returns that example object}.
But how would we create an instance of the (not yet defined) \st{Price} class?
What would we like to assert about it?
What should the API of the \st{Price} class look like?

It is true that TDD forces us to ask such questions, so in this way it \emph{drives} the design process.
\emph{However}, TDD does not otherwise help us in answering these questions.
We are forced to ``guess first'' and try to specify an interface before we can implement anything.

Instead, with EDD, we can exploit the fact that we have examples to iterative and incrementally specify the test (\ie the example method) and \emph{prototype} the object it tests.
In this way, EDD helps us to make small steps towards both specifying the test example method and implementing the code that will make the test (example) green.

\begin{figure}[h]
  \includegraphics[width=\columnwidth]{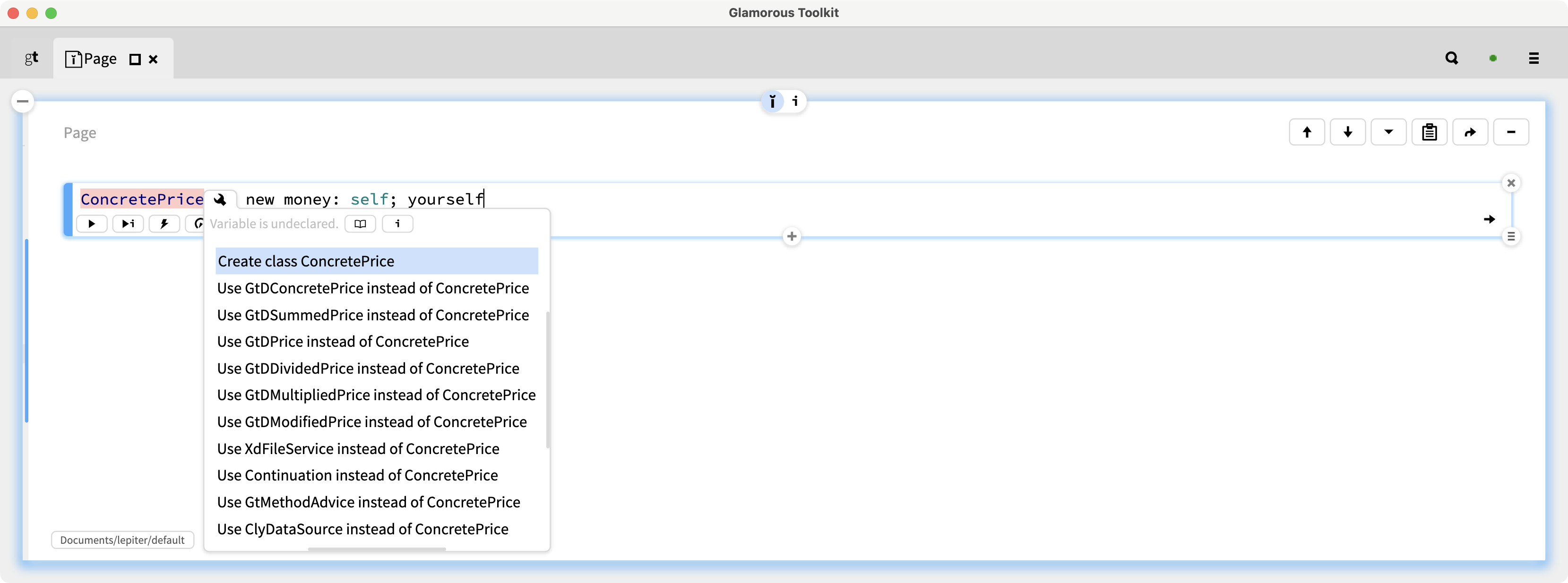}
	\caption{Creating Price object using a fixit dialogue.}
  \label{fig:fixit}
\end{figure}

We can see this illustrated in \autoref{fig:fixit} where we start by writing a snippet that creates a new \lst{ConcretePrice} object with \st{100 euros} as its \st{money} value.
Since the class does not yet exist, we create it with the help of a \emph{fixit dialogue}, and give it a \st{money} slot (instance variable) and setter method.

\begin{figure}[h]
  \includegraphics[width=\columnwidth]{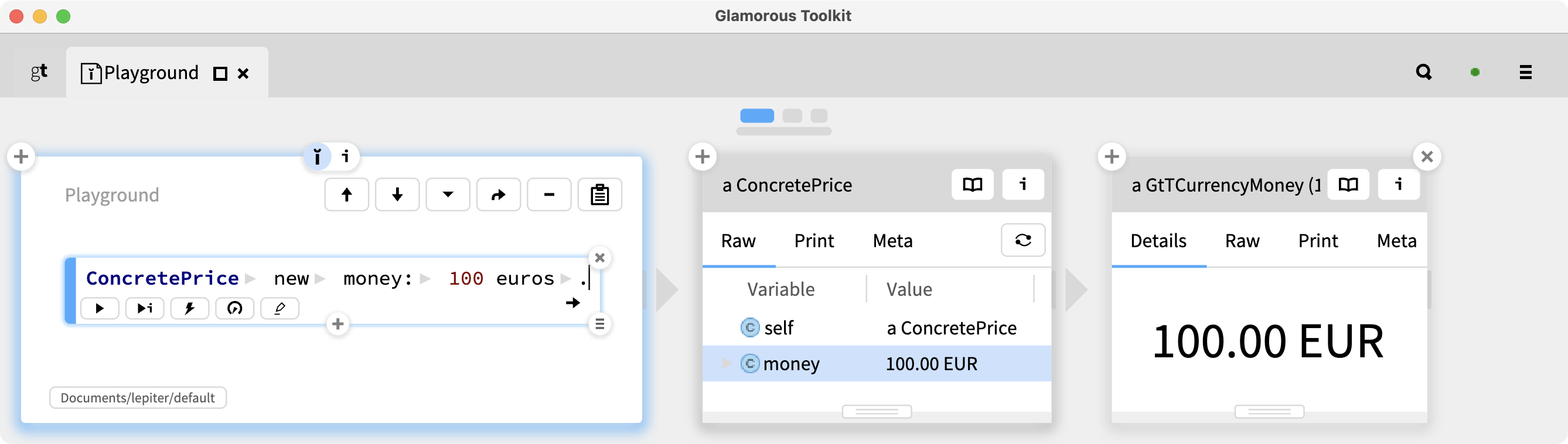}
	\caption{Prototyping a raw Price object.}
  \label{fig:exampleCreationA}
\end{figure}

Once the class is created, we can inspect the result of the expression (middle pane of \autoref{fig:exampleCreationA}), showing that its \st{money} slot is properly initialized.
We can also inspect the slot's value (rightmost pane).

\begin{figure}[h]
  \includegraphics[width=\columnwidth]{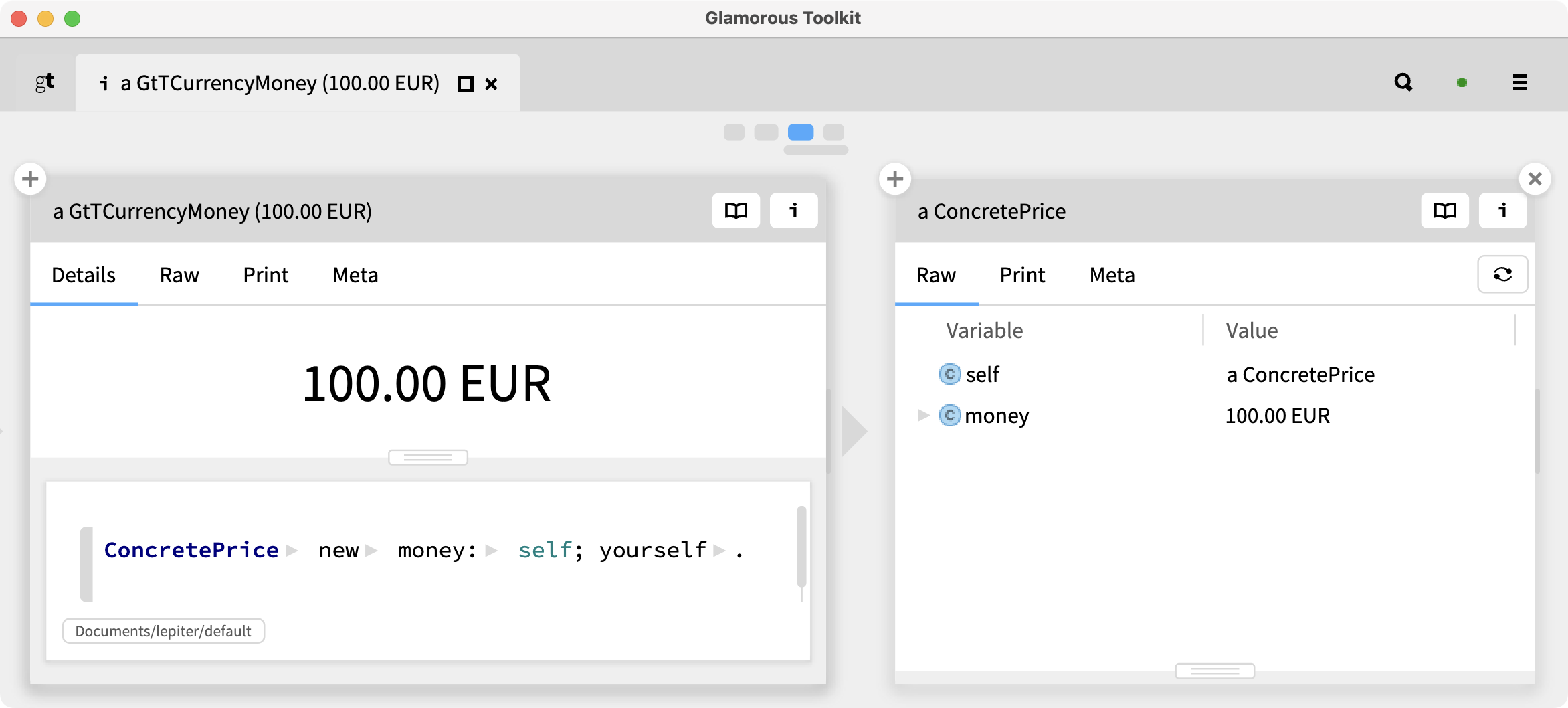}
	\caption{Prototyping a factory method.}
  \label{fig:exampleCreationB}
\end{figure}

At this point we realize that it would be nice to have a factory method to create a \st{Price} from a \st{Money} instance.
In the context of the live \st{100 euro} \st{Money} object (\autoref{fig:exampleCreationB} bottom of left pane) we prototype the code to create a price from this instance.
The code is similar to what we started with, except that the initialization argument is now \st{self}, the \st{Money} instance we are inspecting:
\begin{code}
ConcretePrice new money: self; yourself.
\end{code}
Evaluating this snippet and inspecting the result (right pane) gives us the \st{ConcretePrice} object that we expect (right pane).

\begin{figure}[h]
  \includegraphics[width=\columnwidth]{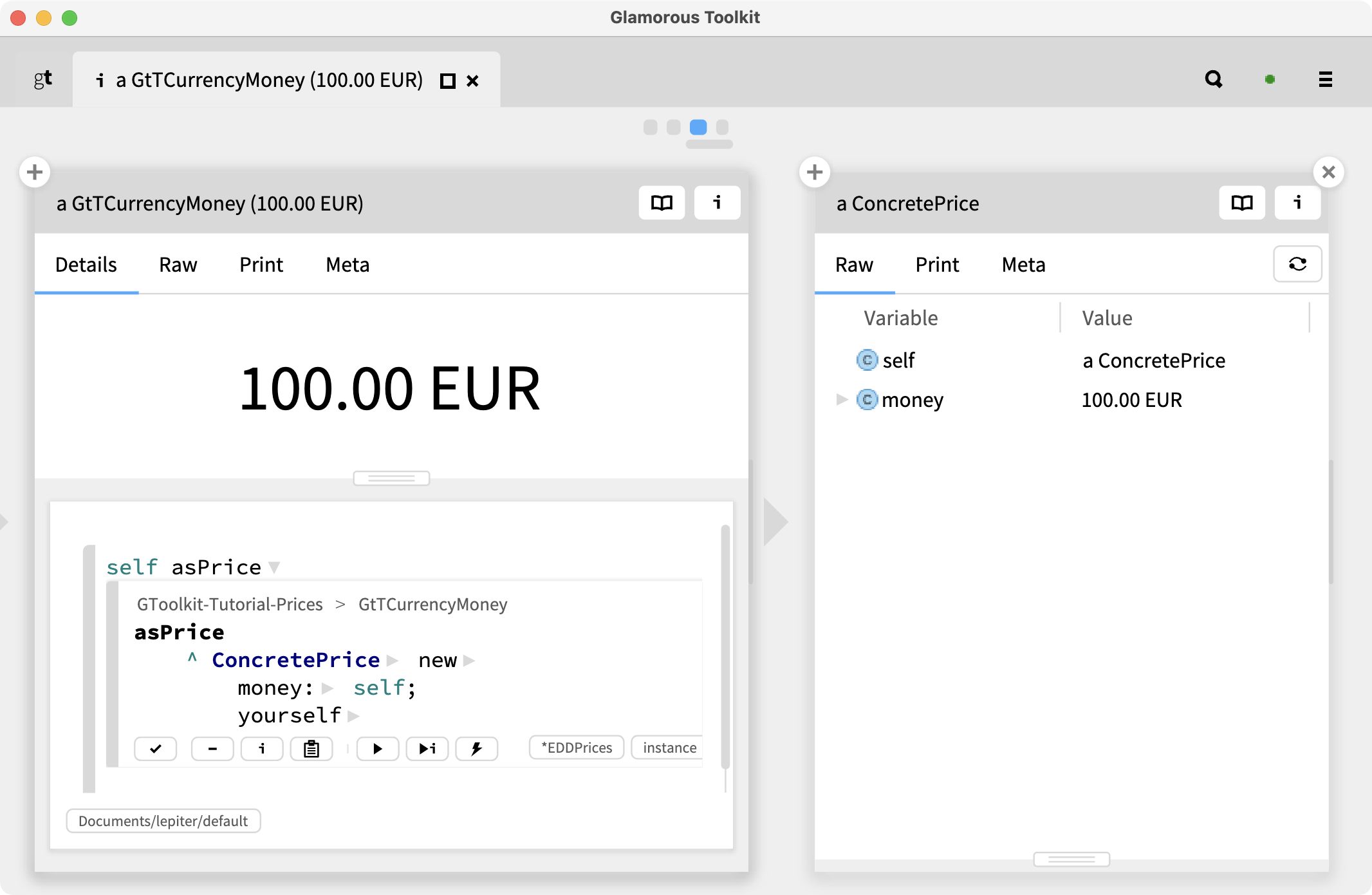}
	\caption{Extracting a factory method.}
  \label{fig:exampleCreationC}
\end{figure}

Now that we have prototyped the factory code, we can extract it as an extension method of the \st{Money} class called \st{asPrice} using an \emph{Extract method} refactoring.
We see the refactored code (\autoref{fig:exampleCreationC}, left pane) as a ``code bubble''~\cite{Brag10a} by expanding \st{asPrice}.
We also see that the refactored snippet still yields the result we want (right pane).

\begin{figure}[h]
  \includegraphics[width=\columnwidth]{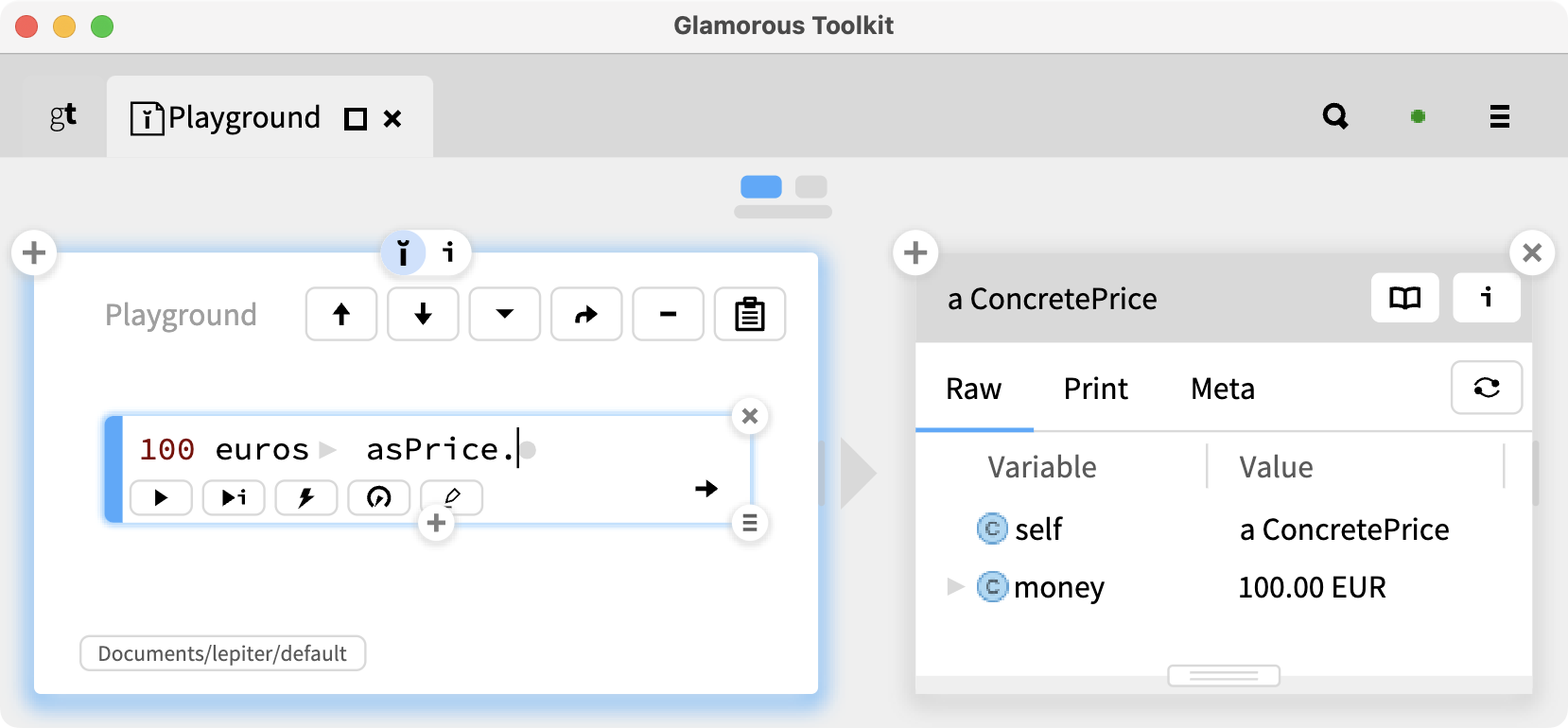}
	\caption{Rewriting the initial snippet.}
  \label{fig:exampleCreationD}
\end{figure}

Now we can go back to our original snippet (\autoref{fig:exampleCreationD}, left pane) and rewrite it as:
\begin{code}
100 euros asPrice.
\end{code}

\begin{figure}[h]
  \includegraphics[width=\columnwidth]{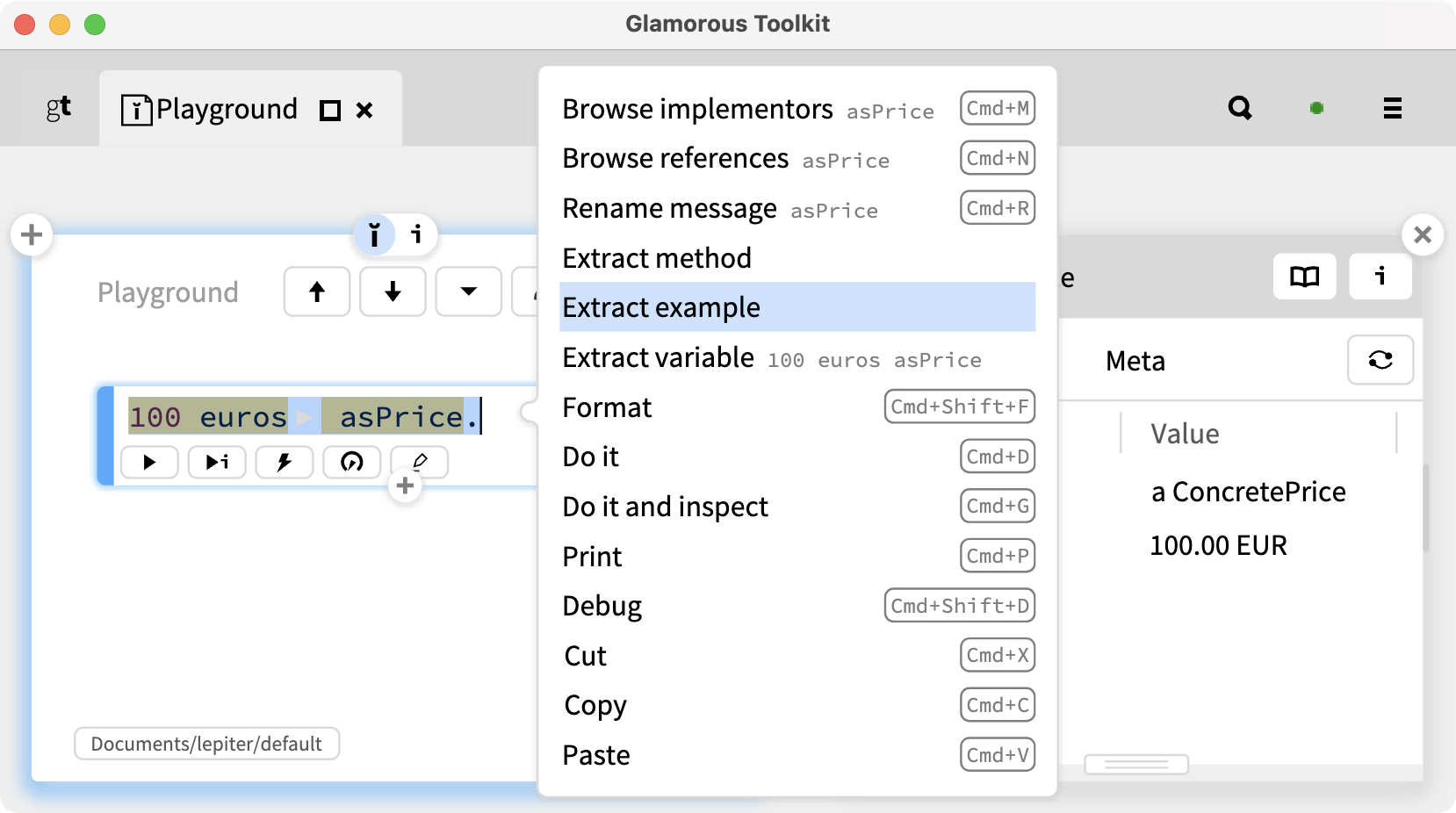}
	\caption{Extracting an example method.}
  \label{fig:exampleExtractionA}
\end{figure}

\begin{figure}[h]
  \includegraphics[width=\columnwidth]{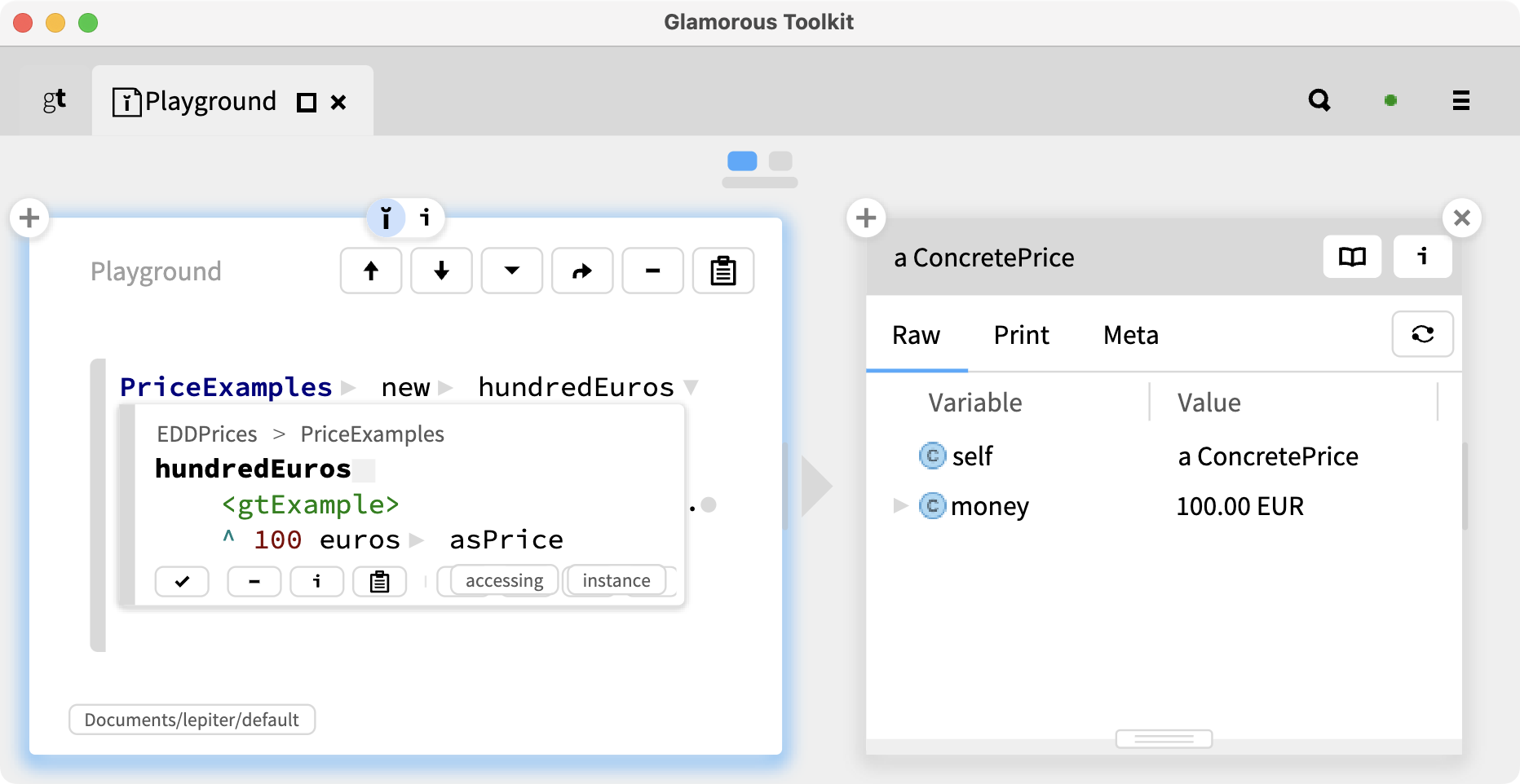}
	\caption{The extracted example method.}
  \label{fig:exampleExtractionB}
\end{figure}

Now we have a nice snippet that creates an example that interests us.
In \autoref{fig:exampleExtractionA} we apply an \emph{Extract example} refactoring to create a new \st{PriceExamples} class with a \st{hundredEuros} example  method.
It is simply a method with a \st{<gtExample>} annotation (analogous to Java method annotations) that flags it as an example method, and which returns the object of interest (\autoref{fig:exampleExtractionB}, left pane with code bubble).
Our example method still does not test anything, so let's prototype that too.
We would expect our \st{hundredEuros} example to be equal to another instance that is created in the same way.

\begin{figure}[h]
  \includegraphics[width=\columnwidth]{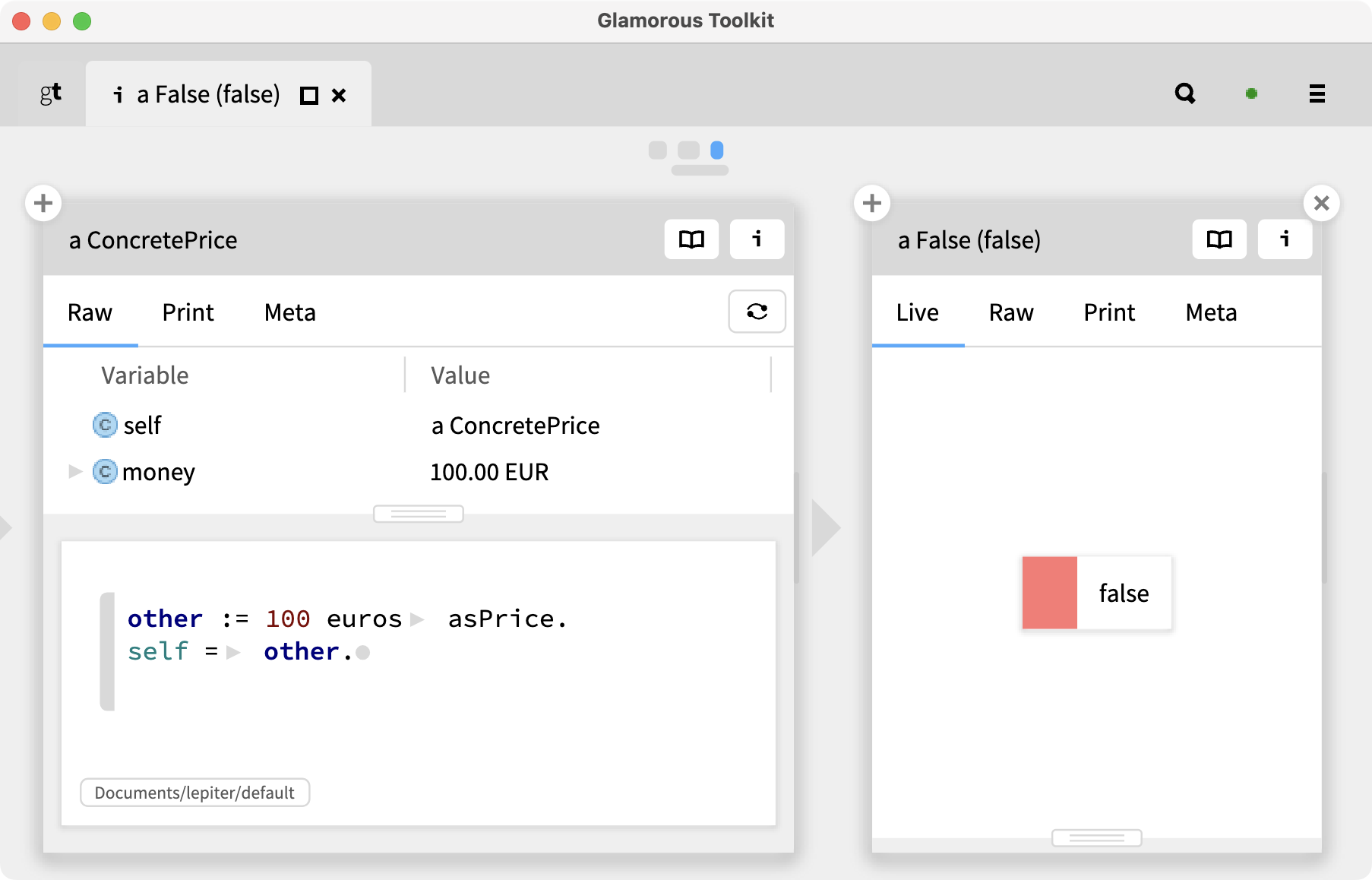}
	\caption{Prototyping an assertion.}
  \label{fig:exampleExtractionC}
\end{figure}

Within the context of the live example (\autoref{fig:exampleExtractionC}, left pane) we prototype the assertion that this example object (\ie \st{self}) is equal to another object created the same way.
Unfortunately this fails (right pane) because our new \st{ConcretePrice} object has not implemented equality, so equality defaults to object identity.

\begin{figure}[h]
  \includegraphics[width=\columnwidth]{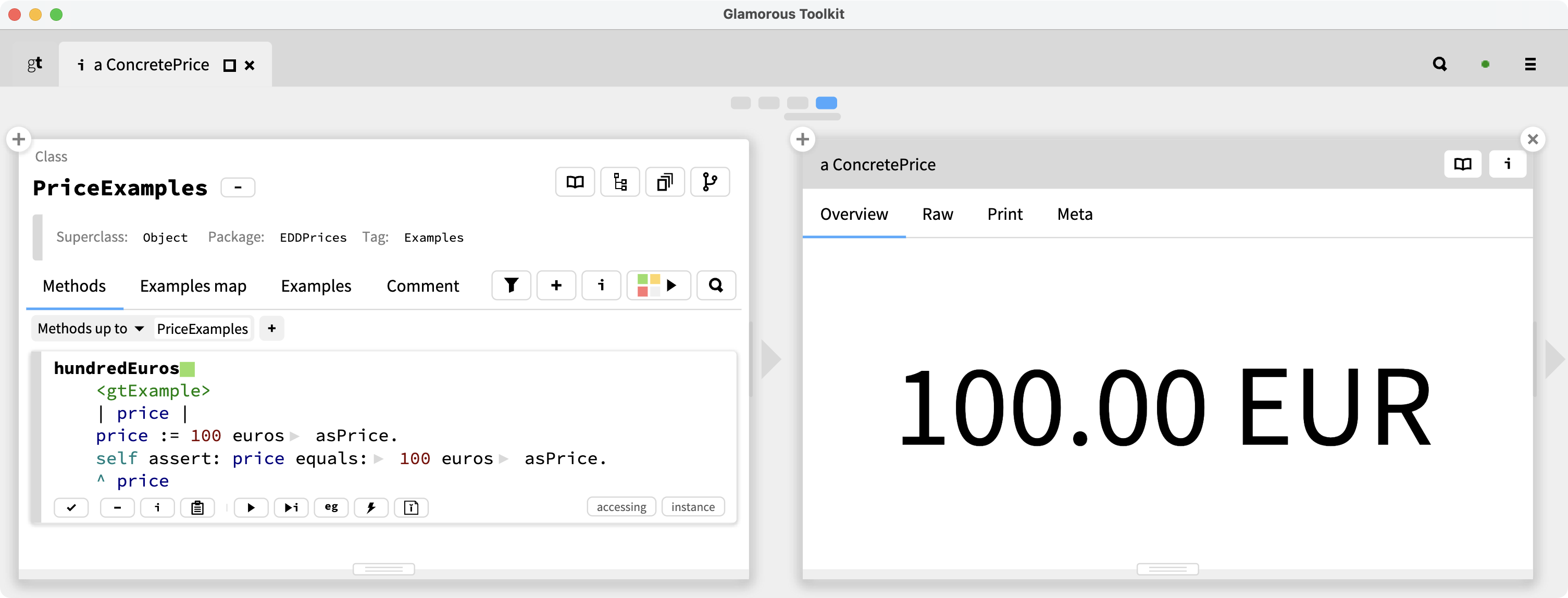}
	\caption{Adding the assertion to the example method.}
  \label{fig:exampleExtractionD}
\end{figure}

We implement the missing method, and now update our example method (\autoref{fig:exampleExtractionD} left pane) with the new assertion.
If we evaluate this example method, it is not only green (left), but also returns an example we can explore (right).

\section{Moldable Examples}\label{sec:moldable}

When we inspect our \st{hundredEuro} example (\autoref{fig:exampleExtractionD}), instead of the original ``raw'' inspector view we see a new \emph{Overview} view showing the value of the concrete price.
How did this happen?
Actually there is a step missing that we will now explain.

\emph{Moldable development} is an approach to constructing \emph{explainable software systems} by augmenting the objects of the software system with dozens of tiny analysis tools that can answer questions about the system.
It can be understood as a refinement of EDD in which objects (examples) are enhanced with custom tools during the development process.

\begin{figure}[h]
  \includegraphics[width=\columnwidth]{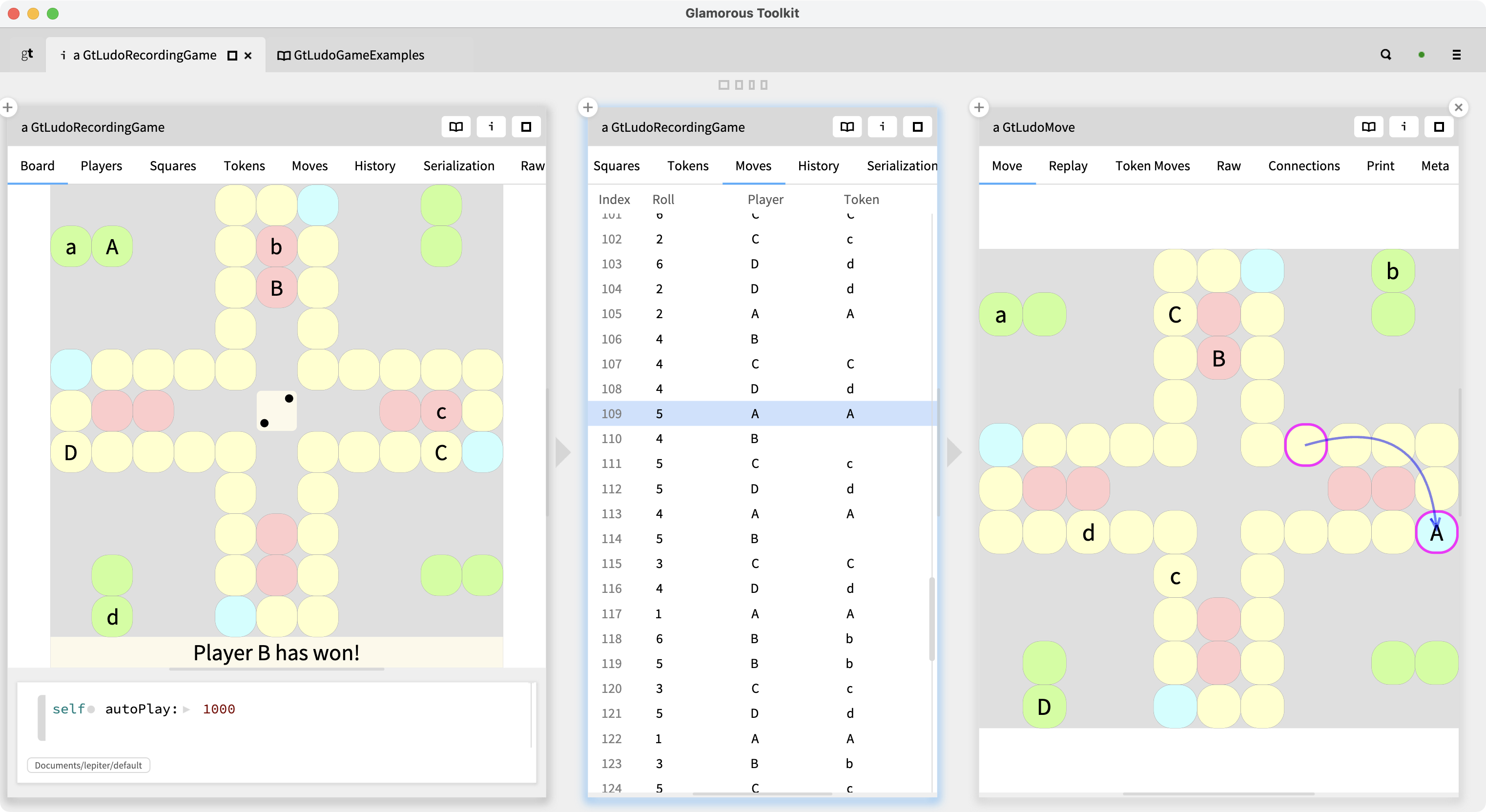}
  \caption{Custom views of a Ludo game.}
  \label{fig:ludoViews}
\end{figure}

Moldable development is made possible with the help of \emph{moldable tools}~\cite{Chis17a}, such as code browsers, debuggers, and object inspectors, that can adapt themselves to the run-time context of an application to enable these analysis tools.
For example, consider the screenshot of a Ludo\footnote{\href{https://web.archive.org/web/20240530004250/https://en.m.wikipedia.org/wiki/Ludo}{https://en.m.wikipedia.org/wiki/Ludo}} game in figure \autoref{fig:ludoViews}.
At the left we see in an object inspector a GUI \emph{Board} view of a running instance of the game that has terminated with player B winning.
In the middle we see a \emph{Moves} view of the same instance, showing us all the past moves leading to the concluding state of the game.
Finally, at the right we inspect a particular move, showing us how the game state was updated in move \#109.

These views have each been created with a few lines of code, in the first and last cases leveraging the existing GUI view of the Ludo game.
The object inspector recognizes that the Ludo game object is an instance of the \lst{GtLudoRecordingGame} class, 
which has been extended with several custom views defined as annotated methods of that class.
Similarly the move object is an instance of the \st{GtLudoMove} class, which has been extended with other views specific to moves.

Two other common types of custom tools are \emph{custom actions} (\eg buttons), which perform a task and possibly spawn another tool such as an inspector, a code editor or an external web browser, and \emph{custom searches}, which query the running object model, and spawn a tool such as an object inspector on the result.

Example methods serve as both the input and output of moldable development.
Typically we start with a ``raw'', unenhanced example, such as we see in \autoref{fig:exampleCreationB}: the \st{ConcretePrice} inspector view shows just a basic ``raw'' view of the instance state of the example.
As we elaborate the examples in the EDD process, we mold them with custom tools that validate the requirements expressed by the examples.
The output of the process is then a molded example that not only checks assertions about its expected behavior, but also exposes that behavior through custom tools.

\begin{figure}[h]
  \includegraphics[width=\columnwidth]{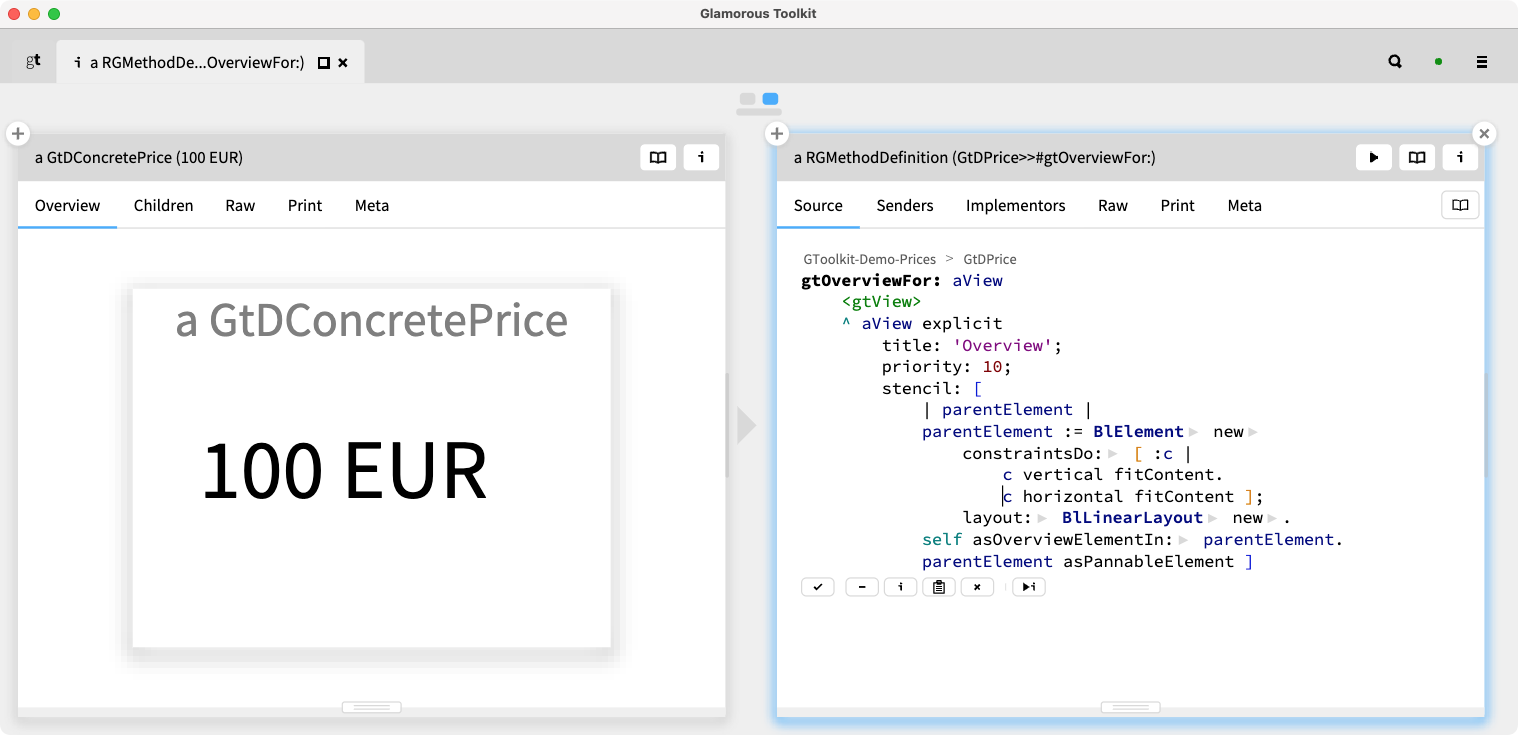}
  \caption{A ConcretePrice object with a custom view.}
  \label{fig:ConcretePrice}
\end{figure}

In \autoref{fig:ConcretePrice} we see another demo version of our \lst{ConcretePrice} class called \st{GtDConcretePrice}.
It has a custom view called \emph{Overview} (left pane) showing that the price is just a fixed amount of money.
In the right pane we see the code that implements this custom view.
The details of the implementation are not important here, but please note that this method consists of just a dozen lines or so of boilerplate code.
Most custom views in GT are short, and the average size of these methods is about a dozen lines.

\begin{figure}[h]
  \includegraphics[width=\columnwidth]{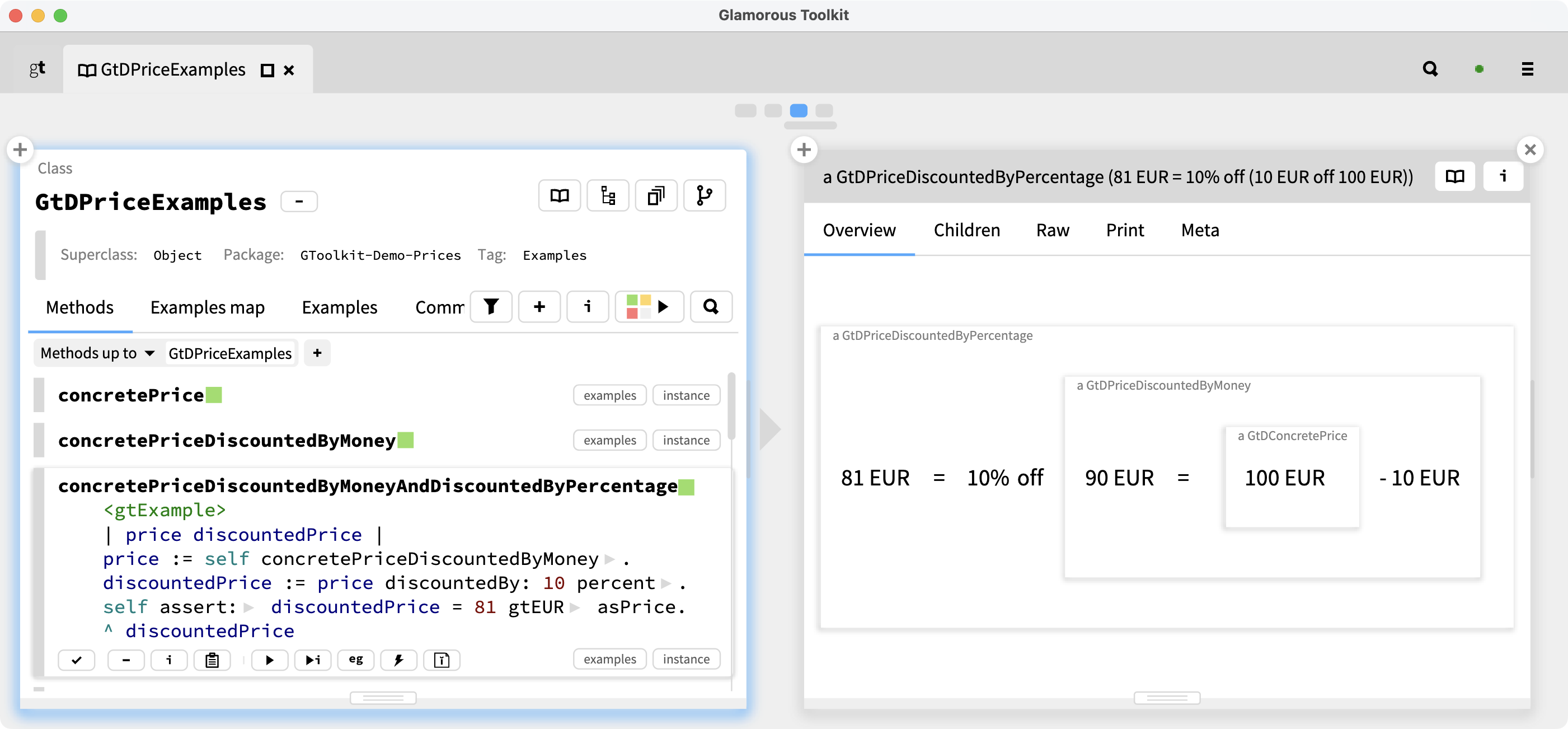}
  \caption{A DiscountedPrice showing how it is composed.}
  \label{fig:DiscountedPrice}
\end{figure}

In \autoref{fig:DiscountedPrice} we see a more complex example of a price that has been discounted twice, first by a fixed amount, and then by a percentage.
In this case, the \emph{Overview} explains how the price has been calculated as a composition of discounts.

\section{Making Systems Explainable with Examples}\label{sec:explainable}

%



Perhaps the most compelling use of examples is within live documentation.
GT includes support for knowledge bases consisting of linked notebook pages that are composed of various kinds of snippets: formatted text, images, code in various programming languages, and live examples.
An example snippet identifies an example method to be evaluated, and a view to be rendered when the notebook page is loaded.

This simple feature enables the creation of various kinds of live documentation, such as live project diaries, interactive tutorials, and live API documentation.

Let's see a few examples from the GT book, the knowledge base that comes bundled with the GT download.

\subsection{Explaining a game}
\begin{figure}[h]
  \includegraphics[width=\columnwidth]{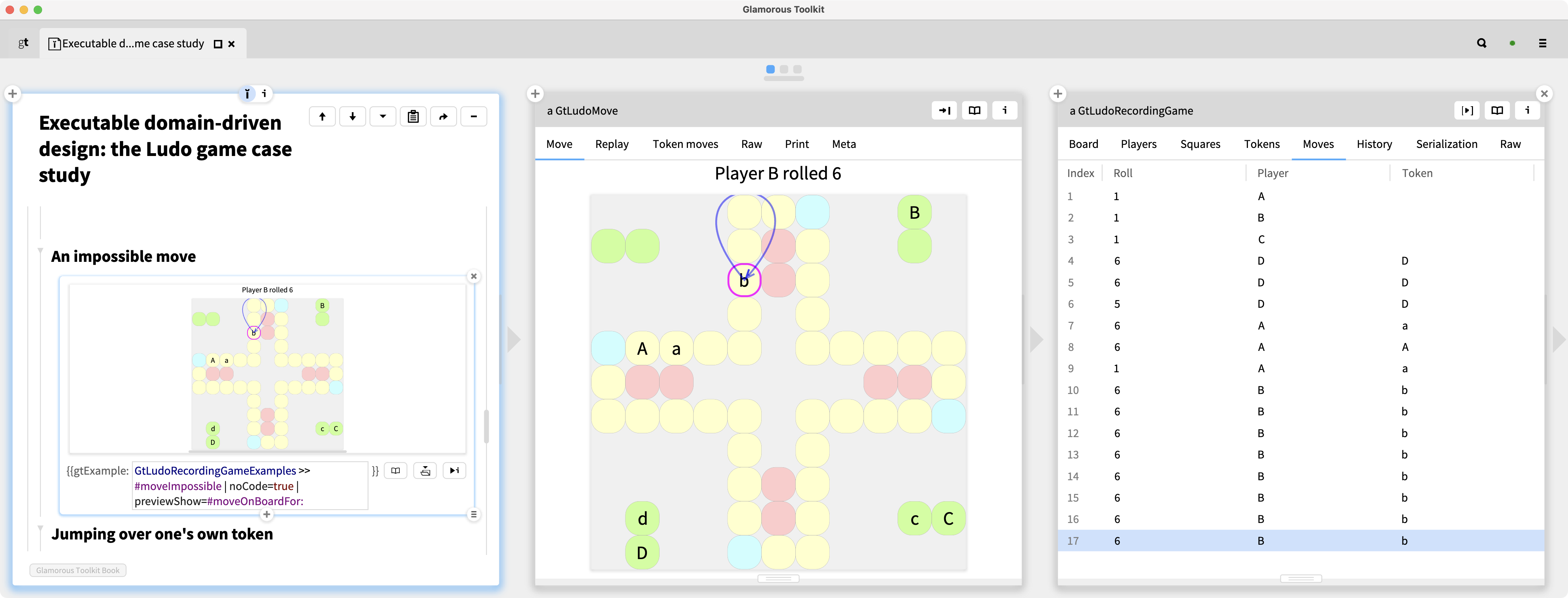}
  \caption{Explaining the rules for Ludo moves.}
  \label{fig:LudoBook}
\end{figure}

In \autoref{fig:LudoBook} we see an extract of a page describing the Ludo game (leftmost pane).
This section of the page describes the rules of the game using examples that illustrate various kinds of moves, in this case an ``impossible move.''
What we see is a custom view of a move object produced by the \st{moveImpossible} example method of the \st{GtLudoRecordingGameExamples} class, illustrating a move that is not possible because there are not enough squares left in the game, thus sending the token back to its current square.
We can dive into the live example (middle pane) and navigate to the replayed game (rightmost pane) to explore the moves leading to this position.

\subsection{Explaining an interpreter implementation}
\begin{figure}[h]
  \includegraphics[width=\columnwidth]{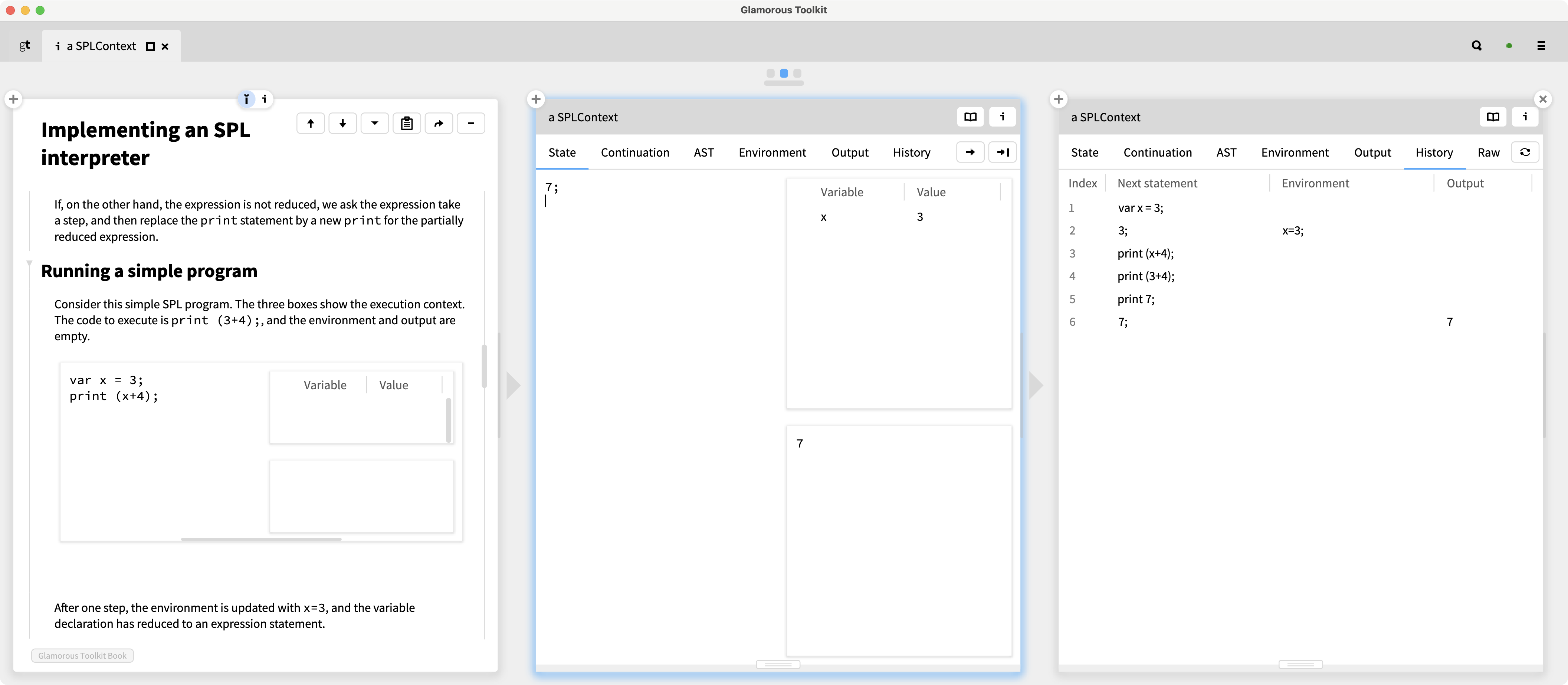}
  \caption{Documenting an interpreter.}
  \label{fig:SPL}
\end{figure}

In \autoref{fig:SPL} we see a notebook page (leftmost page) describing the implementation of a simple programming language (called SPL) using small-step semantic transformations over the abstract syntax tree.
The page contains a live embedded example of simple program.
We can inspect the example (middle pane), execute several steps, and inspect a \st{History} view of the same example (rightmost pane) to see how the environment and output are updated after each transformation step.

\subsection{Explaining a layout algorithm}
\begin{figure}[h]
  \includegraphics[width=\columnwidth]{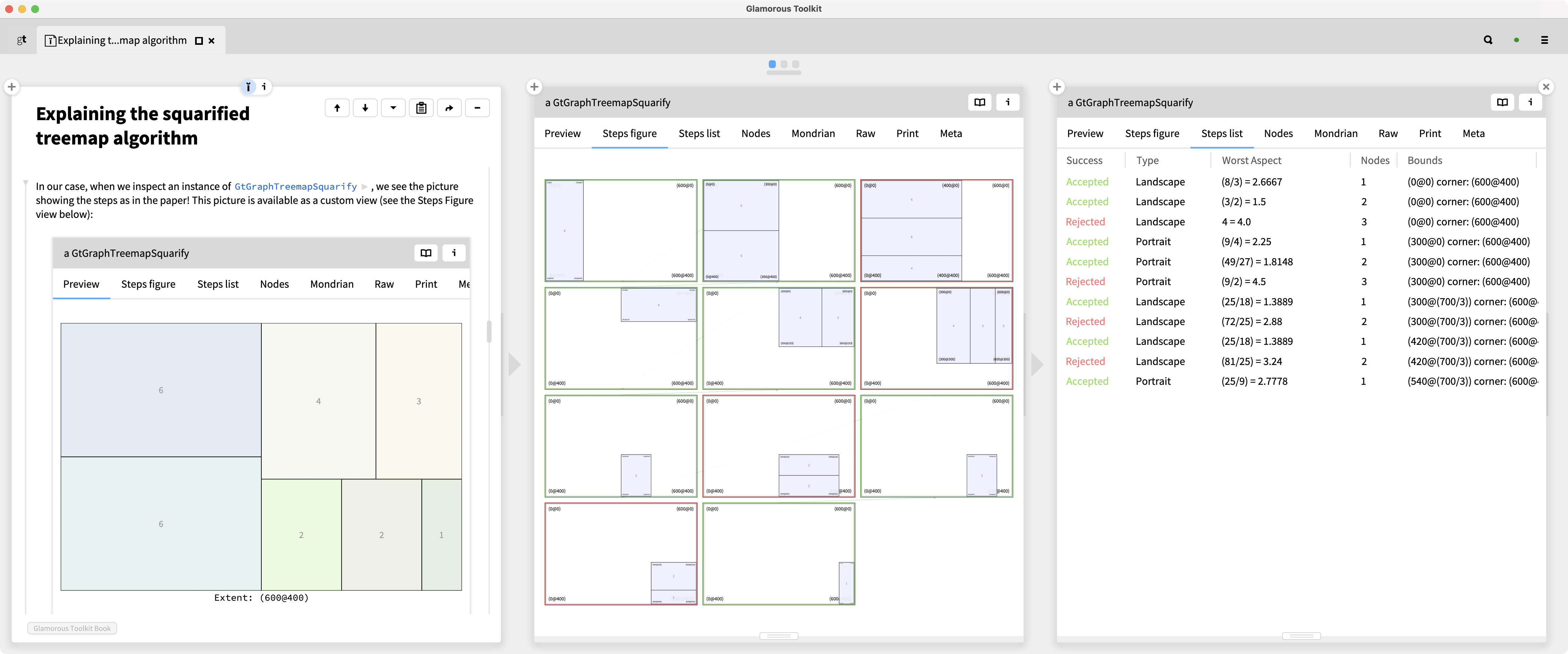}
  \caption{Explaining a squarified TreeMap algorithm.}
  \label{fig:Treemap}
\end{figure}

Examples can also be used effectively to explain algorithms.
In \autoref{fig:Treemap} we see a notebook page (left pane) that explains how a squarified TreeMap layout algorithm works with the help of live examples.
The embedded example not only shows the final layout (left), but by diving into it (middle), we can see the decision steps taken to periodically switch orientation between horizontal and vertical tree map nodes to maintain a ``squarified'' appearance, and we can also see a list view (right) of the steps, explaining how the decisions were made between one possible layout and another.

\subsection{Explaining moldable development}
\begin{figure}[h]
  \includegraphics[width=\columnwidth]{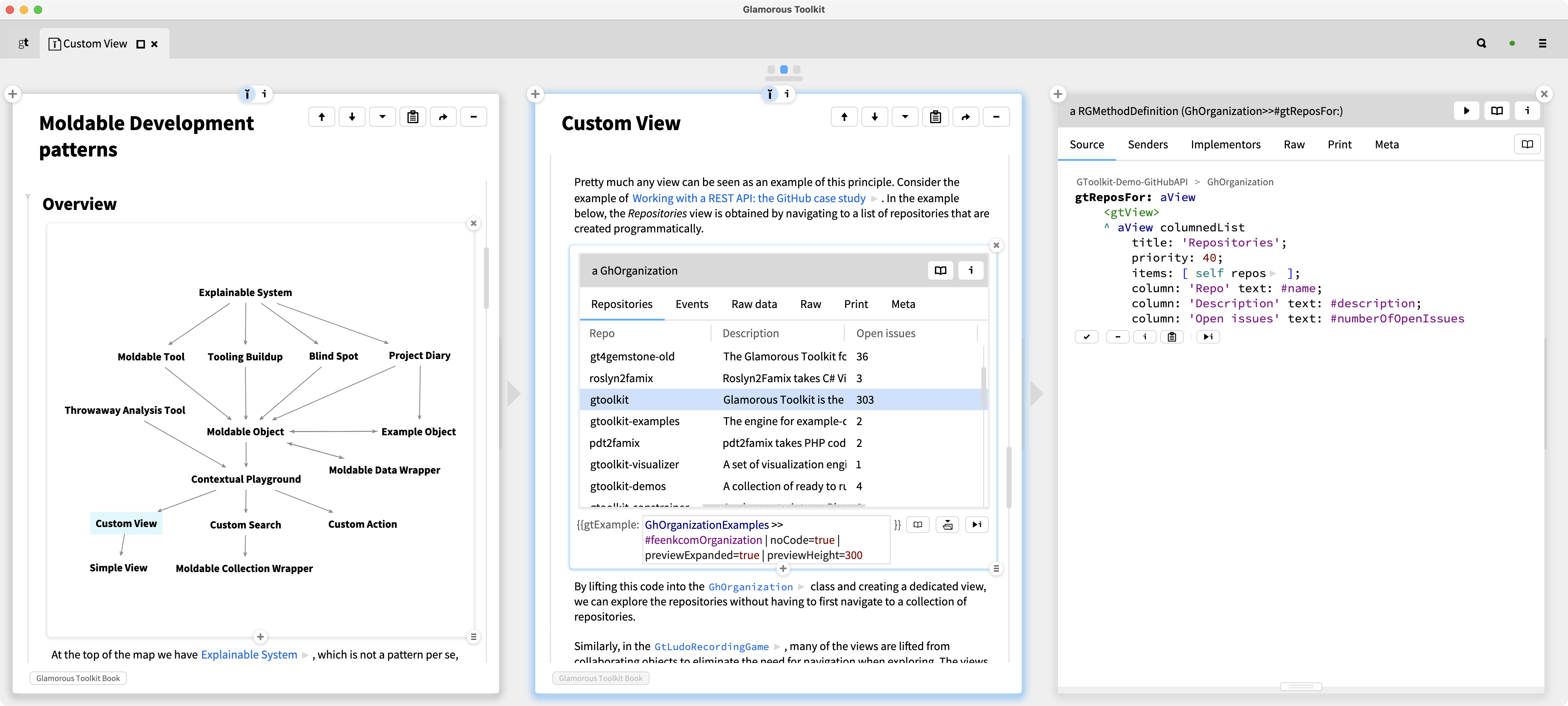}
  \caption{A map of Moldable Development patterns.}
  \label{fig:Patterns}
\end{figure}

Finally, \autoref{fig:Patterns} shows how examples can be used within notebook pages that explain the moldable development process itself in terms of a number of patterns~\cite{Nier24a}.
The \emph{Moldable Development patterns} page (left pane) contains a live embedded map (an example) that allows you to navigate to pages describing the individual patterns.
These pages, such as \emph{Custom View} (middle pane), in turn contain other live examples to illustrate the patterns.
We can interact with the embedded examples, for example, to inspect the source code behind the custom view (right pane).

All of these illustrations show not only how example methods can be used productively to produce live documentation, but that by applying moldable development principles, the live examples can be easily enhanced with lightweight, custom tools that serve to make the examples, and the systems they are part of, truly explainable.


\section{Applying EDD}\label{sec:applying}


How can you apply EDD outside of GT?

First, you need a testing framework in which tests return examples.
JExample~\cite{Kuhn08a}, mentioned earlier, illustrates how this can be done by adapting an existing unit testing framework such as JUnit.
In addition to tests returning objects, it must be possible for tests to reuse existing examples as setups.
The test runners should also be aware that tests return objects, and provide a way to navigate to a live example using an existing object inspector.

Second, to apply EDD as we describe it, it helps to have some support for live programming.
This means having an interactive shell to prototype code that will end up as example methods.
Without this, EDD looks much more like TDD, where you have to ``guess'' the example code without the benefit of iteratively and incrementally prototyping code snippets that will end up being part of the example.

Third, to make use of live examples within documentation, you need a notebook system that will work with live code snippets.
Since examples are generated by simply obtaining the result of evaluating an example method, this should work out of the box, given the existence of a notebook system.

Finally, to really see the benefit of EDD for moldable development, you need the tools of your IDE to be \emph{moldable}~\cite{Chis17a}.
This means that the IDE tools are open to extensions that are dynamically provided by the objects they manipulate, \eg as annotated methods.

Clearly each of these steps entails greater investment in extending the development environment, but it is not necessary to take all the steps at once.
Already the first, relatively simple step of extending the unit testing framework to return examples can bring significant advantages in terms of making green tests explorable and reusable.

\section{Related work}\label{sec:related}



Bush was the first to dream of a computerized ``memex'' of stored knowledge~\cite{Bush45a}, through which a user could trace an associative ``trail'' of interconnected text and multimedia resources.
The memex vision inspired Engelbart's NLS~\cite{Enge68a}, the first system to demonstrate human interaction with a computer mouse, windows, and hypertext features.
Knuth first pioneered the implementation of a computational notebook, called WEB, to support ``literate programming'' through the combination of text, graphics, and live code~\cite{Knut97a}.

Subtext~\cite{Edwa04a} supports \emph{example-centric programming} by placing live examples at the focus of the development process.

Modern notebook systems such as Jupyter\footnote{\url{https://jupyter.org}}, MATLAB Live Scripts\footnote{\url{https://www.mathworks.com/}} and Wolfram Notebooks\footnote{\url{https://www.wolfram.com/notebooks/}} all offer the ability to embed live instances of classes within notebook pages, but these live examples are neither integrated with unit testing frameworks, not do they support custom tooling with the help of moldable tools to tailor the user experience.

Clerk~\cite{Kava23a} is an open-source programming assistant for the Clojure language, which offers moldable, custom views within notebook pages.


The \emph{Test Data Builder} pattern~\cite{Free09a} introduces factory methods for test fixtures, also potentially reducing code duplication in tests, but the resulting objects are only intended as inputs for tests, not their outputs.
The resulting examples are not accessible as the output of a green test.

Cucumber~\cite{Hell17a} is a software tool that supports Behavior-Driven Development through business rules specified in the Gherkin language.
These rules include the specification of ``examples'' (AKA scenarios) that illustrate business rules.
The usage of these examples, however, is restricted to the context of the business rules, and they are not intended as the outputs of tests.

Modern testing frameworks such as pytest~\cite{Okke22a}, allow tests to be parameterized by fixtures specified as separate methods.
Here too, however, fixtures are only seen as inputs to tests, not as outputs.

\section{Conclusion}\label{sec:conclusion}

Having tests be factories for examples is a small, but potentially groundbreaking enhancement.
EDD enhances TDD by making examples rather than tests be the focus of the development process.
By enhancing examples with lightweight, custom tools, they help users answer all kinds of analysis questions.
By embedding examples in live documentation, they make software systems explainable.
As a consequence, up-to-date documentation becomes merely a side-effect of applying EDD.



\bibliographystyle{ACM-Reference-Format}
\bibliography{eddBridge}

\end{document}